\def\year{2023}\relax
\documentclass[letterpaper]{article} % DO NOT CHANGE THIS
\usepackage{aaai22}  % DO NOT CHANGE THIS
\usepackage{times}  % DO NOT CHANGE THIS
\usepackage{helvet}  % DO NOT CHANGE THIS
\usepackage{courier}  % DO NOT CHANGE THIS
\usepackage[hyphens]{url}  % DO NOT CHANGE THIS
\usepackage{graphicx} % DO NOT CHANGE THIS
\usepackage{amsmath}
\usepackage{amsfonts}
\def\UrlFont{\rm}  % DO NOT CHANGE THIS
\usepackage{natbib}  % DO NOT CHANGE THIS AND DO NOT ADD ANY OPTIONS TO IT
\usepackage{caption} % DO NOT CHANGE THIS AND DO NOT ADD ANY OPTIONS TO IT
\DeclareCaptionStyle{ruled}{labelfont=normalfont,labelsep=colon,strut=off} % DO NOT CHANGE THIS
\usepackage{algorithm}
\usepackage{algorithmic}

\usepackage{newfloat}
\usepackage{listings}
\floatstyle{ruled}
\newfloat{listing}{tb}{lst}{}
\floatname{listing}{Listing}
\title{Accelerating Exploration of Marine Cloud Brightening Impacts on Tipping Points Using an AI Implementation of Fluctuation-Dissipation Theorem.}
\author {
    Haruki Hirasawa\textsuperscript{\rm 1},
    Sookyung Kim\textsuperscript{\rm 2},
    Peetak Mitra\textsuperscript{\rm 2,3},
    Subhashis Hazarika\textsuperscript{\rm 2},
    Salva Ruhling-Cachay\textsuperscript{\rm 2,4},
    Dipti Hingmire\textsuperscript{\rm 1},
    Kalai Ramea\textsuperscript{\rm 2},
    Hansi Singh\textsuperscript{\rm 1},
    Philip J. Rasch\textsuperscript{\rm 5}
}
\affiliations {
    \textsuperscript{\rm 1} University of Victoria, Victoria, BC, Canada\\
    \textsuperscript{\rm 2} Palo Alto Research Center, Palo Alto, CA, US\\
    \textsuperscript{\rm 3} Excarta, Palo Alto, CA, US\\
    \textsuperscript{\rm 4} University of California San Diego, San Diego, CA, US\\
    \textsuperscript{\rm 5} University of Washington, Seattle, WA, US\\
    Corresponding authors: hhirasawa@uvic.ca, sookim@parc.com
}

\begin{document}

\maketitle

\begin{abstract}

Marine cloud brightening (MCB) is a proposed climate intervention technology to partially offset greenhouse gas warming and possibly avoid crossing climate tipping points. The impacts of MCB on regional climate are typically estimated using computationally expensive Earth System Model (ESM) simulations, preventing a thorough assessment of the large possibility space of potential MCB interventions. Here, we describe an AI model, named AiBEDO, that can be used to rapidly projects climate responses to forcings via a novel application of the Fluctuation-Dissipation Theorem (FDT). AiBEDO is a Multilayer Perceptron (MLP) model that uses maps monthly-mean radiation anomalies to surface climate anomalies at a range of time lags. By leveraging a large existing dataset of ESM simulations containing internal climate noise, we use AiBEDO to construct an FDT operator that successfully projects climate responses to MCB forcing, when evaluated against ESM simulations. We propose that AiBEDO-FDT can be used to optimize MCB forcing patterns to reduce tipping point risks while minimizing negative side effects in other parts of the climate.
%During training, we apply physics-informed constraints to penalize violations of global climate mass and energy conservation.
\end{abstract}

\section{Introduction}
\subsection{Marine Cloud Brightening}

Tipping points in the climate system are critical components of the climate response to anthropogenic warming, as they have the potential to undergo rapid, self-perpetuating, and possibly irreversible changes \cite{mckay_exceeding_2022}. Should warming approach or cross a threshold that activates such a tipping point, the damage caused by crossing the threshold may be sufficiently severe that a climate intervention ought to be undertaken to prevent it. One such class of interventions are solar radiation modification (SRM) methods which slightly modify the climate's energy budget by scattering away a portion of incoming sunlight (also called solar radiation) to counter some of the effects of greenhouse warming. 

Here, we consider one such SRM technique, Marine Cloud Brightening (MCB), in which sea salt aerosols would be injected into marine boundary layer clouds to increase their albedo \cite{latham_marine_2012}. If MCB were to be deployed with the aim of limiting tipping point risk, it is crucial that we carefully assess if MCB does indeed reduce these risks and rule out MCB scenarios that might cause unintended climate changes or exacerbate tipping points \cite{diamond_assess_2022}. Due to the short atmospheric lifetime of tropospheric aerosol particles, MCB interventions would be highly localized. This presents both a substantial challenge and a potential opportunity, as the possibility space of MCB interventions is vast, both in terms of strength and spatial pattern. Thus, a thorough assessment of the feasibility of MCB interventions must consider a wide range of potential scenarios. On the other hand, it may be possible to find specific patterns of MCB intervention that achieve desirable climate effects while minimizing negative side effects.

Typically, the effect of MCB is evaluated using simulations in Earth System Models (ESMs), which are comprehensive, dynamic models of the coupled atmosphere-ocean-land-ice system \cite{rasch_geoengineering_2009,jones_climate_2009,stjern_response_2018}. However, ESM simulations are computationally expensive, requiring tens of thousands of core-hours to obtain sufficient sample sizes to assess the impact of a given intervention scenario. Thus, they are impractical as tools to explore a wide range of possible MCB intervention patterns. To accelerate this exploration, we have developed AiBEDO, an  AI model that emulates the relationship between atmospheric radiative flux anomalies and resulting surface climate changes. By using AiBEDO to project the climate impact of cloud radiative flux anomalies, we can rapidly evaluate the impact of MCB-like perturbations on the climate.

\subsection{Fluctuation-Dissipation Theorem}

As there are few MCB forcing simulations that have been conducted in the most recent generation of Coupled Model Intercomparison phase 6 (CMIP6) ESMs, we cannot train on an existing repository of ESM responses to MCB forcing. Thus, the design philosophy of AiBEDO borrows from the Fluctuation-Dissipation Theorem (FDT), a theorem emerging from statistical mechanics that posits that the response of a dynamical system to a perturbation can be inferred from the time-lagged correlation statistics of natural internal fluctuations in the system \cite{kubo_fluctuation-dissipation_1966,leith_climate_1975}. Because the climate is such a dynamical system, FDT has been used to estimate the linear response of the climate to a range of forcings: CO$_2$ doubling and solar radiation perturbations \cite{cionni_fluctuation_2004} and regional ocean heat convergence anomalies \cite{liu_sensitivity_2018}, among others.  If the statistics of the dynamical system are Gaussian, the FDT operator $\mathbf{L}$ can be computed by convolving the covariance matrix between the predictor variables $\vec{x}$ and predictand variables $\vec{y}$, $\mathbf{C}_{\vec{y},\vec{x}}(\tau)$, and the autocovariance matrix of $\vec{x}$, $\mathbf{C}_{\vec{x},\vec{x}}(0)$, over time lags $\tau$. The climatological mean response $\left<{\delta \vec{y}}\right> = \left<{y'}\right> - \left<{y}\right>$ (angle brackets indicating the climatological mean) to a constant forcing $\delta \vec{f}$ is then computed as
\begin{equation}
    \left<{\delta \vec{y}}\right> = \mathbf{L}^{-1} \delta \vec{f} = \left[\int_0^\infty \mathbf{C}_{\vec{y},\vec{x}}(\tau) \mathbf{C}_{\vec{x},\vec{x}}^{-1}(0) d\tau \right] \delta\vec{f}
    \label{eq:fdt_linear}
\end{equation}
As FDT is limited to the linear component of the climate response, we seek to use an AI model with the intention of capturing both linear and non-linear components of the response and loosening some of the conditions required by classical FDT \citep[namely that the probability density function of the relevant climate statistics must be Gaussian or quasi-Gaussian; see][]{cionni_fluctuation_2004,majda_high_2010}. 

Here, we define an AiBEDO operator $A_\tau(\vec{x}_i(t))$, which maps the statistical relationship from a given input $\vec{x}_i(t)$ field to an output $\overline{\vec{y}_i(t+\tau)}$ field after some time lag $\tau$, 
\begin{equation}
 A_\tau(\vec{x}_i(t)) ~ : ~ \vec{x}_i(t) ~ \rightarrow ~ \overline{\vec{y}_i(t + \tau)} ~.
\label{eq:aibedo_emulation}
\end{equation}
 with $i$ indexing the different initial conditions sampled from internal climate noise. Due to uncertainties in the initial condition $\vec{x}_i$ (from monthly averaging and discrete sampling of the fields) and the chaotic dynamics of the system, there is no unique mapping from a given input $\vec{x}_i(t)$ to a later output $\vec{y}_i(t+\tau)$. Rather, AiBEDO projects the mean of the distribution of possible $\vec{y}_{i}(t+\tau)$ trajectories after $\tau$ months given the initial conditions $\vec{x}_i(t)$. We denote this mean using an overline, and the output of AiBEDO as $\overline{\vec{y}_i(t+\tau)}$. If we consider a case where $\vec{x}_i(t)$ is perturbed by a infinitesimally small one-month forcing $\vec{\delta f'}$, the mean evolution becomes $A_\tau(\vec{x}_i + \vec{\delta f'}) = \overline{\vec{y}'_i(t+\tau)}$. Linearizing the response, we approximate the effect of the forcing as
%Thus, we can approximate the change in the climate system evolution due to  for a given initial condition $\vec{x_i}$ by linearizing the response:
\begin{equation}
    \begin{split}
    \overline{\delta \vec{y}_i(t+\tau)} &= \overline{\vec{y}_i'(t+\tau)} - \overline{\vec{y}_i(t+\tau)} \\
            &= A_\tau(\vec{x}_i(t) + \vec{\delta f'}) - A_\tau(\vec{x}_i(t))
    \end{split}
    \label{eq:fdt_aibedo}
\end{equation}

We assume that the time-mean climate response is equivalent to the mean response across many different initial conditions $\vec{x}_i$ (ergodicity). Thus, we can compute the climate mean lag-$\tau$ response to a time varying forcing $\delta \vec{f}(t)$ by averaging over $N$ samples of internal variability $\vec{x}_i$. Following FDT, we then integrate the average lag-$\tau$ responses from $\tau=0$ to some upper limit $\tau = T_{max}$, where the response to a perturbation approximately converges to noise (we choose 48 months), to obtain the climate mean response:
\begin{equation}
    \left<{\delta \vec{y}(t)}\right> = \sum_{\tau=0}^{T_{max}} \frac{1}{N}\sum_{i=0}^N{\left(A_{\tau}(\vec{x}_i + \delta \vec{f}(t-\tau)) - A_{\tau}(\vec{x}_i)\right)}
    \label{eq:fdt_aibedo}
\end{equation}
This allows us to replace the linear response function of classic FDT with a non-linear AiBEDO response function. Note, we assume there are no non-linearities between the AiBEDO responses at different lags (i.e., that the effect of $\delta\vec{f}(t)$ is not affected by the changes induced by $\delta\vec{f}(t-1)$, $\delta\vec{f}(t-2)$, etc).

In this study, we discuss the model architecture and training data used to construct this novel AI-based approach to FDT, evaluate the performance of AiBEDO when emulating climate noise, and present a comparison of the AiBEDO response to MCB-like perturbations to the responses in the fully-coupled ESM. Finally, we propose strategies for estimating uncertainties in the AiBEDO response and exploring the possibility space of MCB intervention scenarios using AiBEDO. By assessing a wide range of MCB scenarios on a scale not possible with ESM experimentation, we aim to determine optimal scenarios to avoid crossing potential tipping points and rule out scenarios with undesirable impacts on tipping points.

\section{Methods}
\subsection{Model Architecture}
%%%%%%%%%%%%%%%%%%%
%\begin{figure}
%\begin{center}
%\includegraphics[width=7.5cm]{model_figure.png}
%\end{center}
%\caption{Schematic view of AiBEDO framework}
%\label{fig:aibedo}
%\end{figure}
%%%%%%%%%%%%%%%%%%%%%%
Here, we describe the generation of the AiBEDO operator $A_{\tau}$ to map input radiative flux anomalies at time $t$ (input: $\vec{x}_i(t) \in \mathbb{R}^{d \times c_{in}}$ ) to corresponding output surface climate variable anomalies after a time lag $\tau$ (output: $\overline{\vec{y}_i(t + \tau)} \in \mathbb{R}^{d \times c_{out}}$). To tackle this, we formulate the problem as a pixel-wise regression problem, learning a mapping from input fields to output fields, $A_{\tau}: \mathbb{R}^{d \times c_{in}} \rightarrow \mathbb{R}^{d \times c_{out}}$, where $d$ is the dimension of the data, and $c_{in}$  and $c_{out}$ are input and output channels, respectively, that are comprised of climate variables (listed in Table \ref{tab:variables}). To train the model $\emph{A}_{\tau}$, we minimize $L_{mse}$, the pixel-wise mean squared loss between the estimated climate response output $\widehat{\vec{y}}$ and the ground-truth climate response $\vec{y}$, averaged over all dimensions of output:
\begin{equation}
  \mathcal{L}_{mse}=\frac{1}{c_{out}d} \sum  \|\widehat{ y}_{t+l} -  y_{t+l}\|_2^2,
\end{equation}

%In addition, we minimize the loss for physical constraints on the output, $\mathcal{L}_{phy}$, to penalize unphysical results from the model (see section ``Physical Constraints"). The overall loss $\mathcal{L}$ is defined as a weighted ($\alpha$) sum of $\mathcal{L}_{mse}$ and $\mathcal{L}_{phy}$, which we minimize end-to-end. That is:
%\begin{equation}
%\mathcal{L} = \mathcal{L}_{mse} +  \alpha\mathcal{L}_{phy},
%\end{equation}
%[More to be said here?  How is $\alpha$ determined?]

%Overall schematic of AiBEDO framework is shown in Figure~\ref{fig:aibedo}.

%We minimize the pixel-wise mean squared loss between the estimated climate response output $\widehat{\delta y}_{t+l}$ and the ground-truth climate response $\delta y_{t+l}$:
%\begin{equation}
%  \min_{\Theta} \frac{1}{T} \sum_{t=1}^{T} \frac{1}{dc_{out}}  \|\widehat{\delta y}_{t+l} - \delta y_{t+l}\|_2^2,
%\end{equation}
%where $\Theta$ is a set of parameters in our models $A^l$.
\subsubsection{Spherical Sampling}
The ESM data we use here is originally on a regular latitude-longitude grid, which is difficult to utilize for training purposes due to the large differences in grid areas between points near the equator versus those at the poles. Specifically, it is challenging to accurately depict the Earth's rotational symmetry through the use of two-dimensional meshes, leading to inaccurate representations of significant climate patterns in ML models that assume a two-dimensional format of data. For this reason, we utilize a geodesy-aware spherical sampling that converts the 2D latitude longitude grid to a spherical icosahedral mesh. Icosahedral grids are specified at the lowest resolution by defining twenty equilateral triangles to form a convex polygon, called an icosahedron. The vertices of the icosahedron are equally spaced points on the sphere that circumscribes it. The resolution of the mesh can be increased by dividing each edge of the icosahedron in half and projecting each new point onto the circumscribed sphere. By resampling in this manner, we are able to iteratively increase the resolution on the sphere. Here, we perform bilinear interpolation  (non-conservative) from 2-D climate data to a level-5 icosahedral grid which whose vertices define a 1-D vector of length 10242 (i.e, $d$ =10242) with a nominal resolution of $\sim$220 km.

\subsubsection{Machine Learning method}
\begin{figure}[h]
\begin{center}
\includegraphics[width=7.5cm]{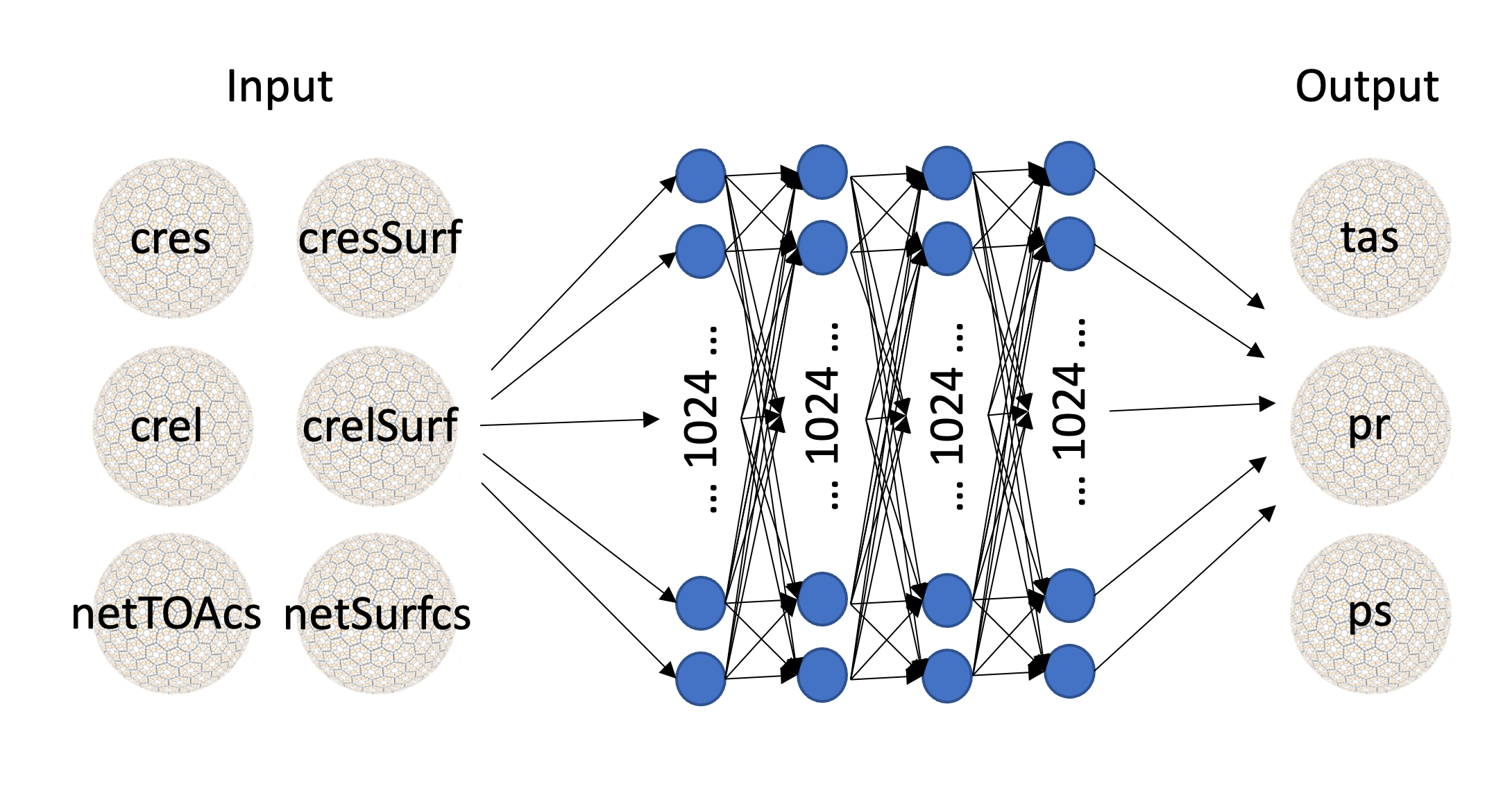}
\end{center}
%\caption{Schematic view of three machine learning methods utilized for AiBEDO model, $A_{\tau}$: (a) Multilayer Perceptrons (MLPs), (b) U-Net, (c) Adaptive Fourier Neural Operator (AFNO) }
\caption{Schematic view of the Spherical Multi-Layer Perceptron (S-MLP) model used in this study.}
\label{fig:ml}
\end{figure}

In this work, we utilize a Multi-Layer Perceptron (MLP) model. MLP models have proven to be effective for spatio-temporal modeling of ESM data \cite{park2019machine, wang2014novel}. MLP is a representative structure of Deep Neural Networks (DNNs) in which an input and an output layer are inter-connected with multiple hidden layers. Each node in a given layer is fully connected with all nodes in the previous layer. The connection between any two nodes represents a weighted value that passes through the connection signal between them. A non-linear activation function is used in each node to represent non-linear correlation in the connection between nodes. The operation between consecutive layers is defined as multiplication between nodes in previous layer and corresponding weight parameters, and applying activation function. Here, we use MLP with 4-hidden layers and 1024 nodes in each layer with layer normalization~\cite{ba2016layer}. We use Gaussian error linear units (Gelu) activation in each layer~\cite{hendrycks2016gaussian}. We combine MLP with the spherical sampling approach to create an S-MLP architecture to generate $A_{\tau}$. A schematic of our S-MLP model architecture is shown in Figure~\ref{fig:ml}.
%%%%%%%%%%%%%%%%%%%
%%%%%%%%%%%%%%%%%%%%%%

%\textbf{(1) Multilayer perceptrons (MLP):}

%\linebreak 
%\textbf{(2) U-Net:}
%U-Net is a symmetric U-shaped convolutional neural network for image-to-image prediction, and consists of a encoder-decoder scheme structure. The encoder extracts visual feature from the input by reducing dimensions in every layer, and the decoder increases the dimensions and predict output with same size as input. Ecoder and decoder are connected with long skip-connections allowing high-resolution features from the encoder are combined with the input of the decoder for better localizing visual feature in prediction.
%\linebreak 

%\textbf{(3) Adaptive Fourier Neural Operator (AFNO):}
%Motivated by the recent successes of transformer-based architectures in climate domain, we adopt the ClimFormer~\cite{ruhling_cachay_climformer_2022} employing Adaptive Fourier Neural Operator (AFNO)~\cite{guibas2021efficient} to tackle our problem.
%FourcastNet is a weather forecasting model using the multi-layer transformer architecture employing AFNO inside. The input is first divided into multiple patches which are embedded in a higher dimensional space with larger number of latent channels and corresponding positional embeddings. Then, positional embeddings are formulated as the sequence of tokens. Tokens are spatially mixed using AFNO followed by subsequent mixing of latent channels accordingly. Mixed embeddings are passed by MLP to learn higher level feature. We repeat this process for each transformer layer. After this process, a linear decoder reconstruct the patches from final embedding.  

\subsection{Training data}
Because the signal-to-noise ratio in short-term climate fluctuations is small, FDT requires a large amount of training data. We use a subset of the Community Earth System Model 2 Large Ensemble (CESM2-LE) as a source of internal climate variations \cite{rodgers_ubiquity_2021} (Table \ref{tab:cesm2}), specifically the 50 ensemble members in which historical simulations are forced with smoothed biomass burning emissions between 1997 and 2014. Each of these 50 ensemble members is forced identically, but is initialized with different initial conditions, meaning that individual members differ only in the chaotic fluctuations internal to the climate system. As such, the CESM2-LE is one of the largest data sets of single-ESM CMIP6-generation simulations for training and testing our model, as it provides a total of nearly 100,000 months of data. 

We use a set of six input variables and three output variables. These variables are listed in Table \ref{tab:variables}. The data are preprocessed by subtracting the ensemble mean of the LE at each grid point, month, and year of the historical time series. This removes both the seasonal cycle and long term secular trends in the data, leaving only monthly fluctuations internal to the system. We then bilinearly remap the data from the original 2D latitude-longitude ESM grid to the spherical icosahedral grid for use by the AI model using Climate Data Operators \citep[cdo;][]{schulzweida_cdo_2022}.

\begin{table*}[t]
\centering
%\resizebox{.95\columnwidth}{!}{
\begin{tabular}{|l|l|l|}
    \hline
    \textbf{Variable} & \textbf{Description} & \textbf{Role in AiBEDO} \\
    \hline
    \texttt{cres} & Net TOA shortwave cloud radiative effect & input \\ \hline
    \texttt{crel} & Net TOA longwave cloud radiative effect & input \\ \hline
    \texttt{cresSurf} & Net Surface shortwave cloud radiative effect & input \\ \hline
    \texttt{crelSurf} & Net Surface longwave cloud radiative effect & input \\ \hline
    \texttt{netTOAcs} & Net TOA clear-sky radiative flux & input \\ \hline
    \texttt{netSurfcs} & Net surface clear-sky radiative flux plus all-sky surface heat flux & input \\ \hline
    \texttt{lsMask} & Land fraction & input \\ \hline\hline
    \texttt{ps} & Surface pressure & output \\ \hline
    \texttt{tas} & Surface air temperature & output \\ \hline
    \texttt{pr} & Precipitation & output \\ \hline
%    \texttt{evspsbl} & Upward surface evaporation & constraint \\ \hline
%    \texttt{netSurfRad} & Net all-sky surface radiative & constraint \\ \hline
%    \texttt{netTOARad} & Net all-sky TOA radiative & constraint \\ \hline
%    \texttt{hfss} & Surface sensible heat flux & constraint \\ \hline\hline
\end{tabular}
\caption{Name, description, and use by AiBEDO of variables derived from CESM2 LE historical smoothed biomass burning monthly mean data. Thus, $c_{in} = 7$ channels and $c_{out} = 3$ channels. All radiative and heat fluxes at the surface and top of atmosphere (TOA) are positive down.}
\label{tab:variables}
\end{table*}

\subsection{Validation dataset}
\begin{table*}[t]
\centering
%\resizebox{.95\columnwidth}{!}{
\begin{tabular}{|l|c|c|c|c|}
    \hline
    \textbf{Experiment} &\textbf{Role} & \textbf{Forcing} & \textbf{Time span} & \textbf{N}\\
    \hline 
    Historical LE & training, testing, validation & historical & 1850 - 2015 & 50 \\
    \hline \hline
    Y2000 Control & perturbation & Year 2000 Fixed SST & 1 - 20 & N/A \\
    \hline
    Y2000 MCB Perturbed & perturbation &\begin{tabular}{@{}c@{}} Year 2000 Fixed SST + \\ MCB in NEP, SEP, and SEA\end{tabular} & 1 - 10 & N/A \\
    \hline \hline
    SSP2-4.5 LE & response validation & SSP2-4.5 & 2015 - 2100 & 17 \\
    \hline
    SSP2-4.5 + ALL MCB & response validation & \begin{tabular}{@{}c@{}} SSP2-4.5 + \\ MCB in NEP, SEP, and SEA\end{tabular} & 2015 - 2065 & 3 \\
    \hline
    SSP2-4.5 + NEP & response validation & SSP2-4.5 + MCB in NEP & 2015 - 2065 & 3 \\
    \hline
    SSP2-4.5 + SEP & response validation & SSP2-4.5 + MCB in SEP & 2015 - 2065 & 3 \\
    \hline
    SSP2-4.5 + SEA & response validation & SSP2-4.5 + MCB in SEA & 2015 - 2065 & 3 \\
    \hline
\end{tabular}
\caption{CESM2 simulations used to train and verify AiBEDO. NEP, SEP, SEA denote regions where 600cm$^{-3}$ CDNC MCB forcing is imposed, where NEP - Northeast Pacific (0 to 30N; 150W to 110W), SEP - Southeast Pacific (30S to 0; 110W to 70W), SEA - Southeast Atlantic (0 to 30N; 25W to 15E). Note the fixed SST simulations use constant climatological conditions, so we do not note specific years for these simulations.}
\label{tab:cesm2}
\end{table*}

To validate AiBEDO's ability to plausibly model the climate response to MCB-like perturbations, we compare the AiBEDO response to responses from a novel set of fully dynamic, coupled CESM2 simulations \cite{hirasawa_impact_2023}. These simulations are summarized in Table \ref{tab:cesm2}. MCB forcing is imposed by increasing in-cloud liquid cloud droplet number concentrations to 600cm$^{-3}$ within three selected regions in the northeast Pacific, southeast Pacific, and southeast Atlantic, together and separately in SSP2-4.5 simulations (Shared Socioeconomic Pathway 2 - 4.5Wm$^{-2}$ forcing). The effect of MCB is then calculated by taking the difference between the perturbed simulations and the baseline SSP2-4.5 simulations. In addition to the coupled CESM2 simulations, we have conducted ``fixed-sea surface temperature" (fixed SST) simulations, wherein the MCB-like forcing is imposed in the model with SSTs held to climatological values. These are used to calculate the effective radiative forcing (ERF) due to the MCB forcing \cite{forster_recommendations_2016}. AiBEDO is perturbed ($\delta \vec{f}$) with the annual mean \texttt{cres}, \texttt{crel}, \texttt{cresSurf}, \texttt{crelSurf}, \texttt{netTOAcs}, and \texttt{netSurfcs} anomaly fields from the year-2000 MCB Perturbed minus year-2000 Control simulations. Thus, we can compare AiBEDO and CESM2 responses to the same MCB ERF. Note that it is crucial that $\delta\vec{f}$ is computed using fixed SST simulations, as radiation anomalies computed this way do not include radiative feedbacks, which are considered to be part of the response rather than the forcing. In principle, the effects of these feedbacks are encoded in the mappings AiBEDO has learned. Thus, AiBEDO responses using radiation perturbations computed from coupled CESM2 simulations avoids ``double counting" the effect of the radiative feedbacks.

%\begin{figure}
%\centering
%\includegraphics[width=1\columnwidth]{cesm_mcb_radiation_pert.png} 
%\caption{Annual mean radiation anomalies calculated from Fixed SST simulations and applied as MCB perturbations to AiBEDO.}
%\label{fig:mcb_forcing}
%\end{figure}

In order to calculate the response to the radiative perturbations, we first run AiBEDO on 480 randomly sampled months of preprocessed CESM2 internal variability radiation anomalies to obtain a control ensemble of AiBEDO outputs. Then, we run AiBEDO on the same 480-month sample, but with the MCB radiation perturbations added to the variability, giving us a perturbed ensemble of AiBEDO outputs. The impact of the MCB perturbations is estimated as the difference between the control and perturbed AiBEDO outputs. This is repeated for the different time lags. This methodology ensures that the input anomaly fields in the simulations are not too different from the model training data set. Running AiBEDO with the regional radiation perturbations results in artifacts, as the near-zero anomalies outside the perturbation regions are entirely unlike any fields the model is trained on. 
%Furthermore, we run AiBEDO on a pool of monthly samples to account for any non-linearities between the background internal variability anomalies and the MCB-like perturbations that AiBEDO has learned, which might result in differences in the AiBEDO output anomalies depending on the background input variability.

\subsection{Model training and inference}

The decoupled weight decay regularization optimization method, AdamW~\cite{loshchilov2017decoupled} was utilized to train our model in an iterative manner. The learning rate was initially set to $2 \times 10^{-4}$ and exponentially decayed at a rate of $1 \times 10^{-6}$ per epoch.  We trained the model for 15 epochs with a batch size of 10. Our S-MLP models have $\sim$ 108M trainable parameters, and it takes around 1 minute per single epoch for training. The model inference takes an average of 0.5 seconds per data point to generate a prediction.

\section{Results}
\subsection{Emulation of Climate Noise}
\begin{figure*}[h!]
\centering
\includegraphics[width=0.8\textwidth]{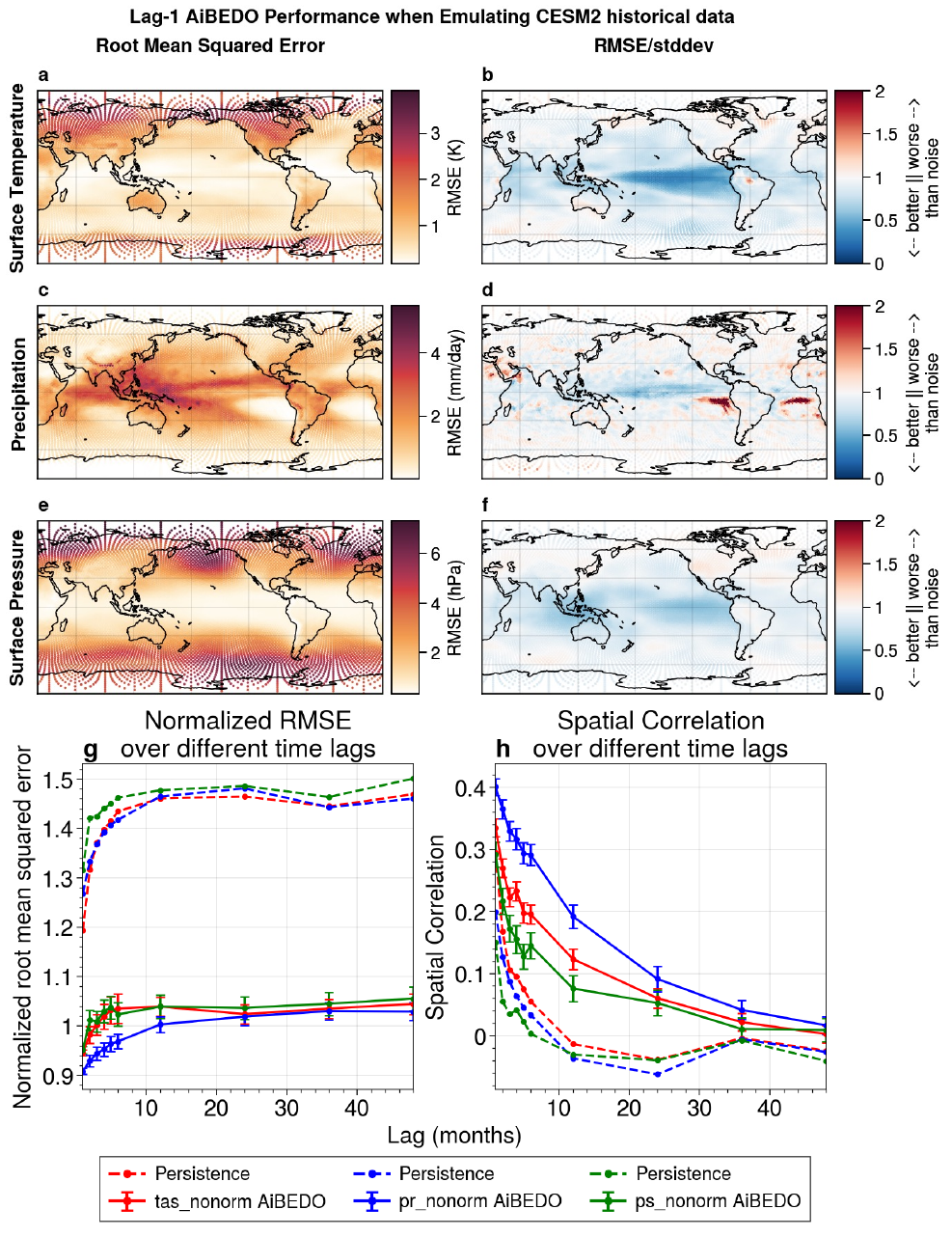} 
\caption{One-month lag AiBEDO compared to CESM2 LE data (a-f). Panels a,c,e show the root mean squared error (RMSE) computed in the time dimension at each icosahedral spherical grid point calculate across 480 months. Panels b,d,f show the ratio of RMSE to the standard deviation of the preprocessed data. Panels a,b show surface temperature (\texttt{tas}), c,d show precipitation (\texttt{pr}), and e,f show surface pressure (\texttt{ps}). Panels h,g show the normalized RMSE (h) and correlations (g) computed along the spatial dimension in solid lines for the three output variables from different AiBEDO lag models. Dashed lines show the normalized RMSE and correlation computed assuming that the anomaly at month 0 remains the same over time (i.e. the persistence null hypothesis). RMSE here is normalized by the climatological spatial standard deviation of the output variable anomalies.}
\label{fig:rmse_cesm2le}
\end{figure*}

We validate the baseline performance of AiBEDO for emulating the connection between input radiative fluxes to the output surface climate variables (i.e. equation \ref{eq:aibedo_emulation}) for a sample of preprocessed CESM2 data in Fig. \ref{fig:rmse_cesm2le} a-f from the CESM2's CMIP6 contribution, data that is not included in the training dataset but uses the same ESM and boundary conditions. This is done by first running AiBEDO with a set of preprocessed input variables from CESM2, then computing the root mean squared error (RMSE) of the resulting AiBEDO output time series with the corresponding lagged CESM2 output time series at each grid point. We find that the RMSE is generally highest in regions where internal variability is also high, such as high \texttt{tas} (Fig. \ref{fig:rmse_cesm2le}a) and \texttt{ps} (Fig. \ref{fig:rmse_cesm2le}e) RMSE at high latitudes and high \texttt{pr} (Fig. \ref{fig:rmse_cesm2le}c) RMSE in the tropics. We then compute the ratio of the RMSE and the CESM2 standard deviation in time: smaller values identify regions where AiBEDO performs best relative to the internal climate noise. This ratio indicates that for all three output variables (Fig. \ref{fig:rmse_cesm2le}b,d,f), AiBEDO performs substantially better in the tropics and subtropics and over oceans, with the tropical Pacific in particular being well represented (this may be a result of the high variance explained by the El Nino-Southern Oscillation). The ratio is slightly under 1 for much of the mid and high latitudes and over land, especially for \texttt{pr}, which may be a consequence of the removal of the seasonal cycle and less direct radiation-surface climate connections in these regions, as they are strongly controlled by synoptic variability. There are a few regions where AiBEDO has a greater than 1 ratio, notably for \texttt{pr} in the dry regions of the subtropical south Pacific and Atlantic \ref{fig:rmse_cesm2le}d. This again may be a consequence of missing seasonal information, as the rainfall in the region is linked to seasonal shifts in the intertropical convergence zone. 

Fig. \ref{fig:rmse_cesm2le} g,h shows the spatial RMSE (normalized by the standard deviation) and correlation scores for different versions of AiBEDO trained at different lags respectively. As lag increases, the predictive skill of the model decreases as expected. Notably, we find that the model outperforms persistence consistently across time lags, indicating AiBEDO has learned a considerable amount of information beyond the simple memory of 0-month temperature anomalies. We see that AiBEDO performs better than background climate noise even at relatively long time 36-month time lags with best performance for precipitation, followed by temperature and surface pressure. Because the normalized RMSE becomes approximately one after 24 months and the correlation drops to zero at 48-months for all three variables, we select 48 months as the upper limit for the time lag integration. 

\subsection{Response to MCB perturbations}

\begin{figure*}[h!]
\centering
\includegraphics[width=0.8\textwidth]{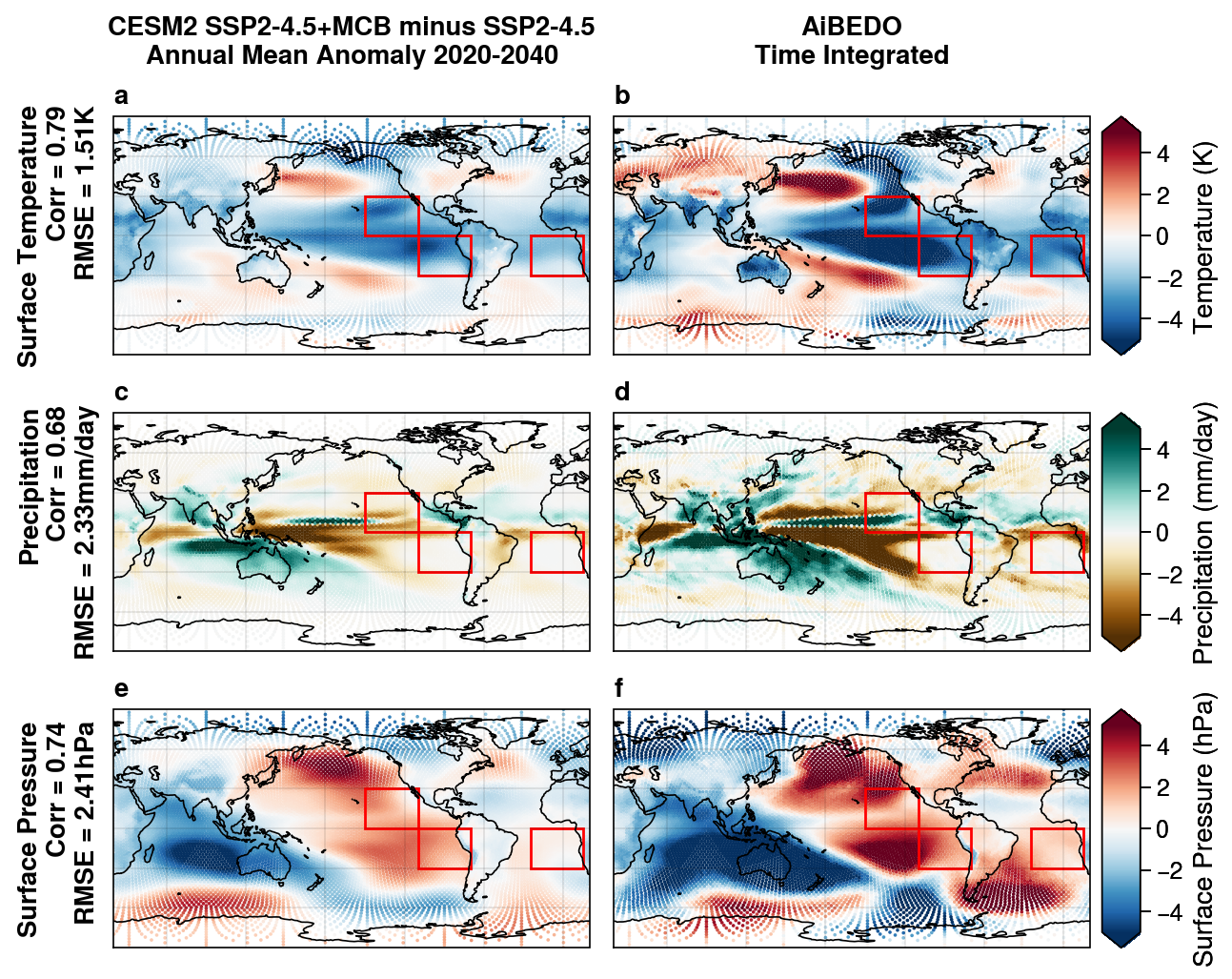} 
\caption{Annual mean temperature (top row - a,b), precipitation (middle row - c,d), and surface pressure (bottom row - e,f) anomalies due to a constant MCB-like forcing for CESM2 (left column) and AiBEDO (right column). Note that the color scale is larger in the AiBEDO figures. Spatial correlation scores and RMSE between the CESM2 and AiBEDO are displayed in the figure labels on the left side.}
\label{fig:cesm_aibedo_mcbresponse}
\end{figure*}

To validate that AiBEDO can plausibly project climate responses to MCB-like perturbations, we compare the CESM2 coupled model responses to those from the lag-integrated AiBEDO responses (i.e. equation \ref{eq:fdt_aibedo}) for radiative flux anomalies computed from fixed-SST MCB simulations. Here, we use a preliminary version of AiBEDO with lags $\tau$ = 1, 2, 3, 4, 5, 6, 12, 24, 36, and 48 months. To compute the lag integral we use Simpson's rule integration to interpolate between the unevenly spaced lags.  Fig. \ref{fig:cesm_aibedo_mcbresponse} shows the CESM2 and AiBEDO responses for the three output variables. We find that AiBEDO is able to reproduce the pattern of climate response to MCB, with correlation scores of 0.83 for \texttt{tas}, 0.72 for \texttt{pr}, and 0.8 for \texttt{ps}. However, there are substantial discrepancies in the magnitude of the responses, with AiBEDO generally projecting larger anomalies than CESM2. This is reflected in the relatively high RMSE when comparing the fields. This magnitude discrepancy may be a result of the missing lags in the integration, which will be filled in future versions of the model.

Nevertheless, AiBEDO successfully identifies key remote teleconnected responses to the MCB forcing, specifically the La Niña-like \texttt{tas} signal in the Pacific, with strong cooling in the tropical Pacific and warming in the midlatitudes east of Asia and Australia, as well as cooling over low-latitude land regions. One notable discrepancy is that northern Eurasia warms in AiBEDO, while there is a weak cooling \texttt{tas} signal in CESM2. This may be in part due to the low signal to noise of the response in this region. AiBEDO also reproduces key \texttt{pr} changes: it projects drying in northeast Brazil, central Africa, and southern North America and Europe and wetting in the Sahel, south and southeast Asia, Australia, and central America. Using these responses, we can estimate the tendency of MCB impacts to affect key regional tipping points. For example, Amazon and Sahel \texttt{pr} changes indicate increased risk of Amazon dieback and Sahel greening, respectively \cite{zemp_self-amplified_2017,mckay_exceeding_2022}. The general cooling of the tropical ocean suggests a reduced risk of coral dieoff tipping points. However, owing to the lower performance of AiBEDO at high latitudes, we may struggle to evaluate key crysopheric tipping points, such as Eurasian and North American permafrost loss.

\begin{figure}
\centering
\includegraphics[width=1\columnwidth]{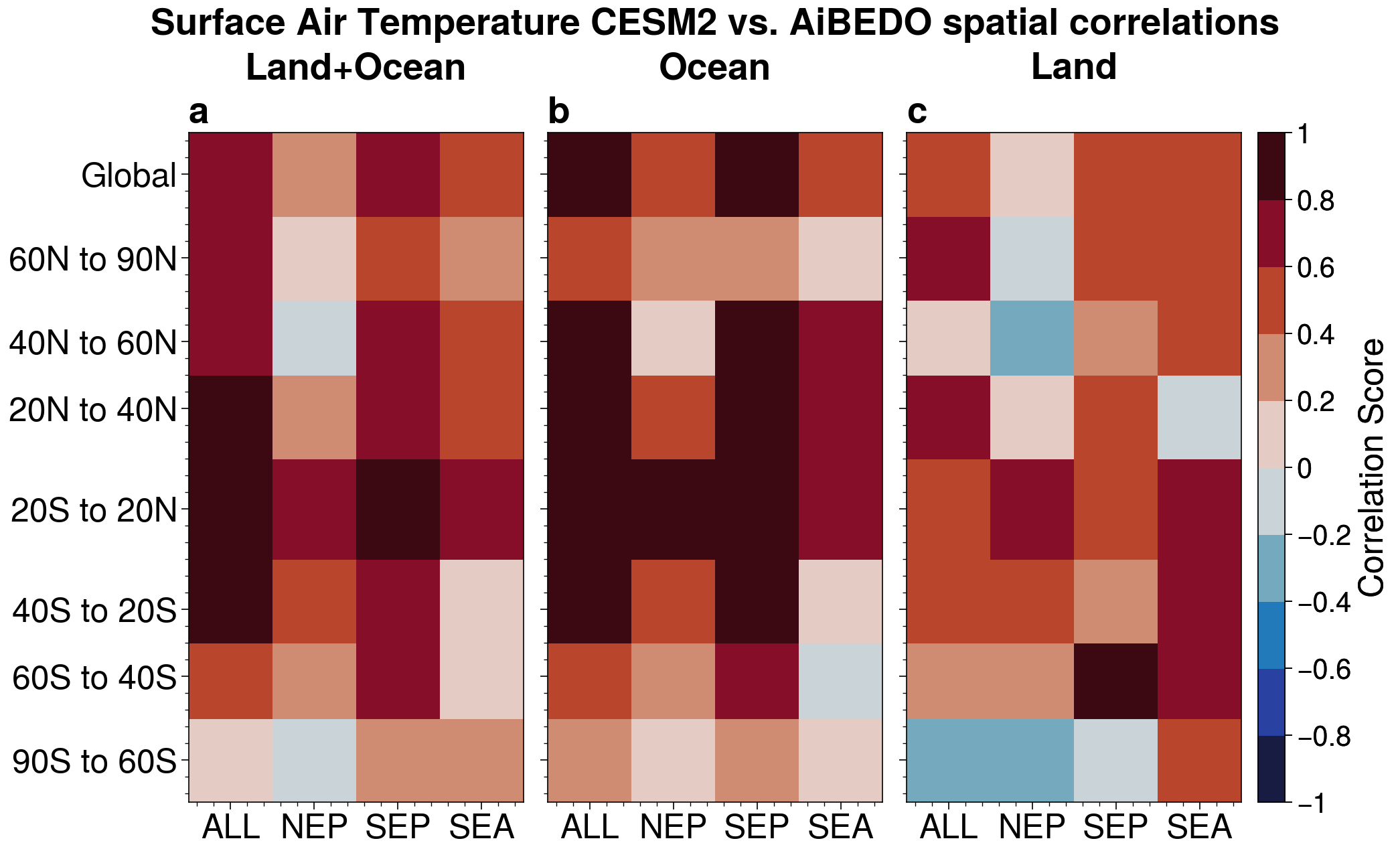} 
\caption{Correlation scores between CESM2 and AiBEDO \texttt{tas} responses to MCB forcing for both land and ocean (a), just ocean (b), and just land (c) in different latitude bands for all MCB regions (ALL), Northeast Pacific (NEP), southeast Pacific (SEP), and Southeast Atlantic (SEA).}
\label{fig:correlation_heatmap}
\end{figure}

We also assess the impact of MCB forcing in the individual NEP, SEP, and SEA regions compared to CESM2 simulations with equivalent regional forcing (Fig. \ref{fig:correlation_heatmap}). We find that AiBEDO performance is weaker when considering these regional perturbations than when all three regions are perturbed together. In particular, AiBEDO's performance when projecting the NEP forcing response declines from a global spatial correlation  of 0.79 for ALL to 0.39 for NEP. AiBEDO correlations scores are better for SEA at 0.48 and best for SEP at 0.75. The weak NEP correlation is due to AiBEDO's too-strong La Niña-like response in the Pacific, possibly indicating that the model over-learns from the El Niño-Southern Oscillation at the expense of other modes of variability. Nevertheless, AiBEDO correctly attributes climate responses to the different forcing regions in several key regions. For example, it correctly identifies that SEP forcing causes La Niña-like cooling and increases in South Asian, West Africa, and Australian rainfall; and finds that SEA forcing causes tropical Pacific warming and Amazon drying (not shown). In all four cases, AiBEDO performs better in the tropics relative to higher latitudes and better over oceans (Fig. \ref{fig:correlation_heatmap}b) than over land (Fig. \ref{fig:correlation_heatmap}c). This aligns with the regions where AiBEDO emulation skill is the highest (Fig. \ref{fig:rmse_cesm2le}b), indicating that the ability of the model to correctly project climate responses to MCB forcing is closely related to its ability to emulate internal variability.
%Aligning with the emulation skill shown in Fig. \ref{fig:rmse_cesm2le}, we find that the correlation between CESM2 and AiBEDO responses is high at low latitudes and over oceans but low at high latitudes and over land. For example, the spatial correlations are above 0.9 for \texttt{tas} and \texttt{ps} over oceans between 20S and 20N \ref{fig:correlation_heatmap}, while they are below 0.4 for all three variables over land between 90S and 60S. Thus, in accordance with the expectations of FDT the ability of the model to correctly project climate responses to MCB forcing is closely related to its ability to emulate internal variability.

\section{Discussion}
In this study, we present a novel framework for rapidly projecting climate responses to forcing by replacing the linear response function in FDT with a non-linear AI model, which we name AiBEDO. AiBEDO is a MLP model with spherical sampling that maps the relationship between monthly-mean radiative flux anomalies and surface climate variable anomalies. The model successfully emulates the connection between variations in radiative fluxes and surface climate variables out to lags of several months. We verify AiBEDO's projections for the case of MCB by comparison to fully coupled CESM2 MCB responses and find that our model is able to skillfully project the pattern of surface temperature, precipitation, and surface pressure response to MCB. We argue the model has sufficient skill to to be useful in estimating the effects of MCB interventions on regional climate indices related to key tipping points, particularly at low latitudes and over oceans. For example, AiBEDO projections reproduce rainfall decreases in the southeast Amazon and increases in the Sahel found in reference CESM2 MCB simulations, indicative of increased risks to tipping points associated with Amazon dieback and Sahel greening. 

To the authors' knowledge, this is the first application of Fluctuation-Dissipation theory to climate data using AI methods. We use a generalization of linear FDT with which we can use a non-linear model to generate mean climate responses to radiative flux anomalies. Notably we use a large single-ESM ensemble of climate model data, which is crucial for AiBEDO to successfully learn the mapping between climate variables, particularly as the time lag increases. Thus, large ensembles like the CESM2-LE are vital resources for training models like AiBEDO. This produces a novel AI model that can plausibly project the impact of MCB on climate, opening the possibility of exploring forcing scenarios on a vastly larger scale than is possible with ESMs. 
%Furthermore, we have developed novel physics-informed constraints to ensure mass and energy conservation on global climate scales, which prevents unrealistic deviations of global mean climate.

We note that while we have selected radiative flux variables as inputs and surface climate variables as outputs here, in principle AI-FDT can be applied to any set of inputs and outputs for which there is sufficient signal-to-noise for a model to learn. Thus, AI models of this kind have the potential to serve as tools with which large existing datasets can be leveraged to generate first look estimates prior to undertaking computationally expensive new ESM simulations, as an AiBEDO projection can be generated in $\mathcal{O}(10^1)$ processor-seconds while just one of the coupled CESM2 MCB simulations we performed here required $\mathcal{O}(10^9)$ processor-seconds.

\subsection{Future Work}
To provide practical information about climate responses to forcing, we must estimate the uncertainty in the projections. Here we only consider the uncertainty due to internal variability in the input data when running AiBEDO, but we must also consider uncertainty due to the underlying training dataset. In particular, because ESMs are only an approximation of the real world, different ESMs exhibit different internal fluctuations. In the case of climate modeling, multi-ESM ensembles, made possible by the Coupled Model Intercomparison Project (CMIP), can be used to quantify this model uncertainty. Thus, we plan to develop an analogous ensemble of AiBEDO models trained on internal fluctuations from different ESMs. Because of the large data requirements of training AiBEDO, we must use single-model initial condition Large Ensembles, of which there exist several from CMIP5 and CMIP6 ESMs \cite{deser_insights_2020}, such as the MPI-ESM1.1 Grand Ensemble \cite{maher_max_2019} and the CanESM2 Large Ensemble \cite{kushner_canadian_2018}. 

Furthermore, though we have verified AiBEDO performance in the response to MCB here (which is largely a shortwave cloud perturbation), AiBEDO includes longwave and clearsky input variables. Thus, AiBEDO may be able to project responses to greenhouse gas and anthropogenic sulphate forcings (both tropospheric pollution and stratospheric injections). We therefore plan to apply AiBEDO to these forcings as well by perturbing the model with ERFs computed from fixed SST simulations with these emissions \cite{forster_recommendations_2016}. 

Using the rapid generation of projections enabled by AiBEDO, we will also develop a method for optimizing MCB forcing patterns to achieve regional climate targets, drawing from the robust existing body of AI-based optimization methods. This will allows us to explore an array of possible MCB scenarios to find which ones may produce desirable regional outcomes. For example, which MCB forcing pattern might achieve the greatest global mean cooling while minimizing drying in the Amazon? Or which patterns minimize polar amplification? This exploration will accelerate the generation of policy-relevant MCB forcing scenarios and allow estimates of the scenario uncertainty in MCB intervention impacts, which is arguably the largest uncertainty in SRM generally \cite{macmartin_scenarios_2022}. 

\section{Acknowledgements}
We thank Linda Hedges of Silver Lining and Brian Dobbins of the National Center for Atmospheric Research for their assistance with AWS and CESM2 computing. The development of AIBEDO is funded under the DARPA AI-assisted Climate Tipping-point Modeling (ACTM) program under award DARPA-PA-21-04-02. AiBEDO training and CESM2 simulations were performed using Amazon Web Services (AWS) computing resources thanks to a generous computing grant from Amazon. We thank the CESM2 Large Ensemble Community Project and supercomputing resources provided by the IBS Center for Climate Physics in South Korea. Documentation of this project can be found at https://aibedo.readthedocs.io/en/latest/index.html.

\bibliography{hirasawa_2023}

\begin{thebibliography}{25}
\providecommand{\natexlab}[1]{#1}

\bibitem[{Ba, Kiros, and Hinton(2016)}]{ba2016layer}
Ba, J.~L.; Kiros, J.~R.; and Hinton, G.~E. 2016.
\newblock Layer normalization.
\newblock \emph{arXiv preprint arXiv:1607.06450}.

\bibitem[{Cionni, Visconti, and Sassi(2004)}]{cionni_fluctuation_2004}
Cionni, I.; Visconti, G.; and Sassi, F. 2004.
\newblock Fluctuation dissipation theorem in a general circulation model: {FDT}
  {IN} {A} {GENERAL} {CIRCULATION} {MODEL}.
\newblock \emph{Geophysical Research Letters}, 31(9): n/a--n/a.

\bibitem[{Deser et~al.(2020)Deser, Lehner, Rodgers, Ault, Delworth, DiNezio,
  Fiore, Frankignoul, Fyfe, Horton, Kay, Knutti, Lovenduski, Marotzke,
  McKinnon, Minobe, Randerson, Screen, Simpson, and Ting}]{deser_insights_2020}
Deser, C.; Lehner, F.; Rodgers, K.~B.; Ault, T.; Delworth, T.~L.; DiNezio,
  P.~N.; Fiore, A.; Frankignoul, C.; Fyfe, J.~C.; Horton, D.~E.; Kay, J.~E.;
  Knutti, R.; Lovenduski, N.~S.; Marotzke, J.; McKinnon, K.~A.; Minobe, S.;
  Randerson, J.; Screen, J.~A.; Simpson, I.~R.; and Ting, M. 2020.
\newblock Insights from {Earth} system model initial-condition large ensembles
  and future prospects.
\newblock \emph{Nature Climate Change}, 10(4): 277--286.

\bibitem[{Diamond et~al.(2022)Diamond, Gettelman, Lebsock, McComiskey, Russell,
  Wood, and Feingold}]{diamond_assess_2022}
Diamond, M.~S.; Gettelman, A.; Lebsock, M.~D.; McComiskey, A.; Russell, L.~M.;
  Wood, R.; and Feingold, G. 2022.
\newblock To assess marine cloud brightening's technical feasibility, we need
  to know what to study—and when to stop.
\newblock \emph{Proceedings of the National Academy of Sciences}, 119(4):
  e2118379119.

\bibitem[{Forster et~al.(2016)Forster, Richardson, Maycock, Smith, Samset,
  Myhre, Andrews, Pincus, and Schulz}]{forster_recommendations_2016}
Forster, P.~M.; Richardson, T.; Maycock, A.~C.; Smith, C.~J.; Samset, B.~H.;
  Myhre, G.; Andrews, T.; Pincus, R.; and Schulz, M. 2016.
\newblock Recommendations for diagnosing effective radiative forcing from
  climate models for {CMIP6}: {RECOMMENDED} {EFFECTIVE} {RADIATIVE} {FORCING}.
\newblock \emph{Journal of Geophysical Research: Atmospheres}, 121(20):
  12,460--12,475.

\bibitem[{Hendrycks and Gimpel(2016)}]{hendrycks2016gaussian}
Hendrycks, D.; and Gimpel, K. 2016.
\newblock Gaussian error linear units (gelus).
\newblock \emph{arXiv preprint arXiv:1606.08415}.

\bibitem[{Hirasawa et~al.(2023)Hirasawa, Hingmire, Singh, Rasch, Kim, Hazarika,
  Mitra, and Ramea}]{hirasawa_impact_2023}
Hirasawa, H.; Hingmire, D.; Singh, H.; Rasch, P.~J.; Kim, S.; Hazarika, S.;
  Mitra, P.; and Ramea, K. 2023.
\newblock Impact of {Regional} {Marine} {Cloud} {Brightening} {Interventions}
  on {Climate} {Tipping} {Points}.
\newblock \emph{In Preparation}.

\bibitem[{Jones, Haywood, and Boucher(2009)}]{jones_climate_2009}
Jones, A.; Haywood, J.; and Boucher, O. 2009.
\newblock Climate impacts of geoengineering marine stratocumulus clouds.
\newblock \emph{Journal of Geophysical Research: Atmospheres}, 114(D10):
  2008JD011450.

\bibitem[{Kubo(1966)}]{kubo_fluctuation-dissipation_1966}
Kubo, R. 1966.
\newblock The fluctuation-dissipation theorem.
\newblock \emph{Reports on Progress in Physics}, 29(1): 255.

\bibitem[{Kushner et~al.(2018)Kushner, Mudryk, Merryfield, Ambadan, Berg,
  Bichet, Brown, Derksen, Déry, Dirkson, Flato, Fletcher, Fyfe, Gillett, Haas,
  Howell, Laliberté, McCusker, Sigmond, Sospedra-Alfonso, Tandon, Thackeray,
  Tremblay, and Zwiers}]{kushner_canadian_2018}
Kushner, P.~J.; Mudryk, L.~R.; Merryfield, W.; Ambadan, J.~T.; Berg, A.;
  Bichet, A.; Brown, R.; Derksen, C.; Déry, S.~J.; Dirkson, A.; Flato, G.;
  Fletcher, C.~G.; Fyfe, J.~C.; Gillett, N.; Haas, C.; Howell, S.; Laliberté,
  F.; McCusker, K.; Sigmond, M.; Sospedra-Alfonso, R.; Tandon, N.~F.;
  Thackeray, C.; Tremblay, B.; and Zwiers, F.~W. 2018.
\newblock Canadian snow and sea ice: assessment of snow, sea ice, and related
  climate processes in {Canada}'s {Earth} system model and climate-prediction
  system.
\newblock \emph{The Cryosphere}, 12(4): 1137--1156.

\bibitem[{Latham et~al.(2012)Latham, Bower, Choularton, Coe, Connolly, Cooper,
  Craft, Foster, Gadian, Galbraith, Iacovides, Johnston, Launder, Leslie,
  Meyer, Neukermans, Ormond, Parkes, Rasch, Rush, Salter, Stevenson, Wang,
  Wang, and Wood}]{latham_marine_2012}
Latham, J.; Bower, K.; Choularton, T.; Coe, H.; Connolly, P.; Cooper, G.;
  Craft, T.; Foster, J.; Gadian, A.; Galbraith, L.; Iacovides, H.; Johnston,
  D.; Launder, B.; Leslie, B.; Meyer, J.; Neukermans, A.; Ormond, B.; Parkes,
  B.; Rasch, P.; Rush, J.; Salter, S.; Stevenson, T.; Wang, H.; Wang, Q.; and
  Wood, R. 2012.
\newblock Marine cloud brightening.
\newblock \emph{Philosophical Transactions of the Royal Society A:
  Mathematical, Physical and Engineering Sciences}, 370(1974): 4217--4262.

\bibitem[{Leith(1975)}]{leith_climate_1975}
Leith, C. 1975.
\newblock Climate {Response} and {Fluctuation} {Dissipation}.
\newblock \emph{Journal of the Atmospheric Sciences}, 32: 2022--2026.

\bibitem[{Liu et~al.(2018)Liu, Lu, Garuba, Leung, Luo, and
  Wan}]{liu_sensitivity_2018}
Liu, F.; Lu, J.; Garuba, O.; Leung, L.~R.; Luo, Y.; and Wan, X. 2018.
\newblock Sensitivity of {Surface} {Temperature} to {Oceanic} {Forcing} via
  q-{Flux} {Green}’s {Function} {Experiments}. {Part} {I}: {Linear}
  {Response} {Function}.
\newblock \emph{Journal of Climate}, 31(9): 3625--3641.

\bibitem[{Loshchilov and Hutter(2017)}]{loshchilov2017decoupled}
Loshchilov, I.; and Hutter, F. 2017.
\newblock Decoupled weight decay regularization.
\newblock \emph{arXiv preprint arXiv:1711.05101}.

\bibitem[{MacMartin et~al.(2022)MacMartin, Visioni, Kravitz, Richter,
  Felgenhauer, Lee, Morrow, Parson, and Sugiyama}]{macmartin_scenarios_2022}
MacMartin, D.~G.; Visioni, D.; Kravitz, B.; Richter, J.; Felgenhauer, T.; Lee,
  W.~R.; Morrow, D.~R.; Parson, E.~A.; and Sugiyama, M. 2022.
\newblock Scenarios for modeling solar radiation modification.
\newblock \emph{Proceedings of the National Academy of Sciences}, 119(33):
  e2202230119.

\bibitem[{Maher et~al.(2019)Maher, Milinski, Suarez‐Gutierrez, Botzet,
  Dobrynin, Kornblueh, Kröger, Takano, Ghosh, Hedemann, Li, Li, Manzini, Notz,
  Putrasahan, Boysen, Claussen, Ilyina, Olonscheck, Raddatz, Stevens, and
  Marotzke}]{maher_max_2019}
Maher, N.; Milinski, S.; Suarez‐Gutierrez, L.; Botzet, M.; Dobrynin, M.;
  Kornblueh, L.; Kröger, J.; Takano, Y.; Ghosh, R.; Hedemann, C.; Li, C.; Li,
  H.; Manzini, E.; Notz, D.; Putrasahan, D.; Boysen, L.; Claussen, M.; Ilyina,
  T.; Olonscheck, D.; Raddatz, T.; Stevens, B.; and Marotzke, J. 2019.
\newblock The {Max} {Planck} {Institute} {Grand} {Ensemble}: {Enabling} the
  {Exploration} of {Climate} {System} {Variability}.
\newblock \emph{Journal of Advances in Modeling Earth Systems}, 11(7):
  2050--2069.

\bibitem[{Majda, Abramov, and Gershgorin(2010)}]{majda_high_2010}
Majda, A.~J.; Abramov, R.; and Gershgorin, B. 2010.
\newblock High skill in low-frequency climate response through fluctuation
  dissipation theorems despite structural instability.
\newblock \emph{Proceedings of the National Academy of Sciences}, 107(2):
  581--586.

\bibitem[{McKay et~al.(2022)McKay, Staal, Abrams, Winkelmann, Sakschewski,
  Loriani, Fetzer, Cornell, Rockström, and Lenton}]{mckay_exceeding_2022}
McKay, D. I.~A.; Staal, A.; Abrams, J.~F.; Winkelmann, R.; Sakschewski, B.;
  Loriani, S.; Fetzer, I.; Cornell, S.~E.; Rockström, J.; and Lenton, T.~M.
  2022.
\newblock Exceeding 1.5°{C} global warming could trigger multiple climate
  tipping points.
\newblock \emph{Science}, 377: 6611.

\bibitem[{Park, Yoo, and Nadiga(2019)}]{park2019machine}
Park, J.~H.; Yoo, S.; and Nadiga, B. 2019.
\newblock Machine learning climate variability.
\newblock In \emph{Proceedings of the 33rd Conference on Neural Information
  Processing Systems (NeurIPS), Vancouver, BC, Canada}, 8--14.

\bibitem[{Rasch, Latham, and Chen(2009)}]{rasch_geoengineering_2009}
Rasch, P.~J.; Latham, J.; and Chen, C.-C.~J. 2009.
\newblock Geoengineering by cloud seeding: influence on sea ice and climate
  system.
\newblock \emph{Environmental Research Letters}, 4(4): 045112.

\bibitem[{Rodgers et~al.(2021)Rodgers, Lee, Rosenbloom, Timmermann,
  Danabasoglu, Deser, Edwards, Kim, Simpson, Stein, Stuecker, Yamaguchi,
  Bódai, Chung, Huang, Kim, Lamarque, Lombardozzi, Wieder, and
  Yeager}]{rodgers_ubiquity_2021}
Rodgers, K.~B.; Lee, S.-S.; Rosenbloom, N.; Timmermann, A.; Danabasoglu, G.;
  Deser, C.; Edwards, J.; Kim, J.-E.; Simpson, I.~R.; Stein, K.; Stuecker,
  M.~F.; Yamaguchi, R.; Bódai, T.; Chung, E.-S.; Huang, L.; Kim, W.~M.;
  Lamarque, J.-F.; Lombardozzi, D.~L.; Wieder, W.~R.; and Yeager, S.~G. 2021.
\newblock Ubiquity of human-induced changes in climate variability.
\newblock \emph{Earth System Dynamics}, 12(4): 1393--1411.

\bibitem[{Schulzweida(2022)}]{schulzweida_cdo_2022}
Schulzweida, U. 2022.
\newblock {CDO} {User} {Guide} (2.1.0).

\bibitem[{Stjern et~al.(2018)Stjern, Muri, Ahlm, Boucher, Cole, Ji, Jones,
  Haywood, Kravitz, Lenton, Moore, Niemeier, Phipps, Schmidt, Watanabe, and
  Kristjánsson}]{stjern_response_2018}
Stjern, C.~W.; Muri, H.; Ahlm, L.; Boucher, O.; Cole, J. N.~S.; Ji, D.; Jones,
  A.; Haywood, J.; Kravitz, B.; Lenton, A.; Moore, J.~C.; Niemeier, U.; Phipps,
  S.~J.; Schmidt, H.; Watanabe, S.; and Kristjánsson, J.~E. 2018.
\newblock Response to marine cloud brightening in a multi-model ensemble.
\newblock \emph{Atmospheric Chemistry and Physics}, 18(2): 621--634.

\bibitem[{Wang et~al.(2014)Wang, Zhang, Wang, Han, and Kong}]{wang2014novel}
Wang, J.; Zhang, W.; Wang, J.; Han, T.; and Kong, L. 2014.
\newblock A novel hybrid approach for wind speed prediction.
\newblock \emph{Information Sciences}, 273: 304--318.

\bibitem[{Zemp et~al.(2017)Zemp, Schleussner, Barbosa, Hirota, Montade,
  Sampaio, Staal, Wang-Erlandsson, and Rammig}]{zemp_self-amplified_2017}
Zemp, D.~C.; Schleussner, C.-F.; Barbosa, H. M.~J.; Hirota, M.; Montade, V.;
  Sampaio, G.; Staal, A.; Wang-Erlandsson, L.; and Rammig, A. 2017.
\newblock Self-amplified {Amazon} forest loss due to vegetation-atmosphere
  feedbacks.
\newblock \emph{Nature Communications}, 8(1): 14681.

\end{thebibliography}

\end{document}